# Certain characteristics of financial management strategies of people living in extreme poverty

Balázs Bazsalya

29.05.2020

Sociologist, Research team member of the 'Financial and Public Services' research project at Corvinus University of Budapest

Abstract

This study presents the structure of financial management of incomes, expenses, and borrowing practices of households in extreme poverty, based on a survey conducted in two disadvantaged regions in Hungary. Additional to that, I shed light on financial management practices of households in extreme poverty based on the analysis of in-depth interviews.

I draw theoretical conclusions and explanations, building upon the experiences of empiric materials about the attitudes and behaviors of financial management of households in extreme poverty, and the reasons and consequences thereof.

JEL codes: E26, I32, Z13

Keywords: social exclusion, debt, extreme poverty, financial management

This research was supported by the Higher Education Institutional Excellence Program of the Ministry of Innovation and Technology in the framework of the 'Financial and Public Services' research project (NKFIH-1163-10/2019) at Corvinus University of Budapest.

**Introduction and theoretical background**

Extreme poverty is not such a concept, that would have an exact definition in the literature. It is used as an everyday expression to describe people who find it hard or even impossible to sustain basic living conditions in a long term. However, describing living conditions might differ in time or by region. Evidently, basic needs were defined differently a few decades ago than what they are now, and those mean something else in an American or European city in comparison to a village in Asia or Africa. Therefore, who is considered poor depends on time and location. Nevertheless, I decided to use this term to describe the characteristics of a group of people, whose behavior, habits, and attitude can be defined by dearth, uncertainty, living from one day to the next, and the lack of mobility options. It does not describe a temporary problem but a permanent state of social exclusion for generations. Experiencing such social exclusion creates deep cultural and habitual differences, that are occasionally hard to decode or interpret from the perspectives of the dominant social norms. Furthermore, these peculiarities are often stigmatized and defined as obstacles of social mobility with the implicit or explicit assumption that poor people are responsible for their situation and if they changed their behavior, they would not be poor (Csoba, 2009). This study - among other things - attempts to present the characteristics of how people living in extreme poverty manage their finances, and give an insight into the conflicts and biases that come from these cultural differences locally.[1]

One may ask about the number of people living in extreme poverty, or what definitions do different statistics apply. Without describing the different measurements[2], I use the most widely applied measurement of Eurostat. I do this because it makes international and timeline-based comparison. The measurements used here are parts of the Laeken indicators set by the European Commission in 2001, to provide reliable and comparable data on poverty and social exclusion in EU member states. Since 2005, new member states of the EU use the same methodology allowed by these numbers as part of the Statistics on Income and Living Conditions - EU-SILC. One of these statistics of the European Statistical Office (Eurostat) that measures poverty is called "at risk of poverty or social exclusion - AROPE".

---

[1] In this study, I focus on the financial management of households. A household's economy or subsistence strategies would include broader concepts, in general management of all resources available, not only the management of finances (Nagy 2003).

[2] About the concept, measurement and dilemmas of poverty see: Havasi-Altorjai 2005 Bánfalvi 2014.

By the definition of Eurostat, AROPE is based on three different subindicators, all of them are suitable for grasping certain aspects of poverty: (1) at risk of poverty: is the share of people with an equivalized disposable income (after social transfer) below the at-risk-of-poverty threshold, which is set at 60 % of the national median equivalized disposable income after social transfers,[3] (2) severely materially deprived shows, that the person cannot afford at least 4 out of the 9 listed items [4], (3) living in a household with a very low work intensity: persons living in households with very low work intensity is defined as the number of persons living in a household where the members of working age worked less than 20% of their total potential during the previous 12 months. The number of people at the risk of poverty or social exclusion has decreased significantly in the past years in Hungary. This number is also decreasing in other member states of the European Union, although to a lesser extent than in Hungary.

---

[3] In Hungary, according to OECD2, the poverty threshold in 2018 was the 60% of the equivalent median income calculated per consumption unit, an amount of 93.313 HUF per month, a higher number than in previous years. In 2010, this threshold was about 61.889 HUF.

[4] The units are the following: (1) to pay their rent, mortgage or utility bills; (2) to keep their home adequately warm; (3) to face unexpected expenses; (4) to eat meat or proteins regularly; (5) to go on holiday; (6) a television set; (7) a washing machine; (8) a car; (9) a telephone. It is important to mention, that another indicator is about to be added, the material and social deprivation rate. This indicator is interpreting the definition of deprivation in a slightly broader sense including social elements as well. In case of the lack of 5 items out of 13 leads to deprivation in this indicator. It uses the following indicators: faceing unexpected expenses; one week annual holiday away from home; avoid arrears (in mortgage, rent, utility bills and/or hire purchase installments); afford a meal with meat, chicken or fish or vegetarian equivalent every second day; keep their home adequately warm; a car/van for personal use; replace worn-out furniture; replace worn-out clothes with some new ones; have two pairs of properly fitting shoes; spend a small amount of money each week on him/herself ("pocket money"); have regular leisure activities; get together with friends/family for a drink/meal at least once a month; have an internet connection. (Source: https://ec.europa.eu/eurostat/web/products-eurostat-news/-/DDN-20171212-1). According to this indicator a significant drop can be seen in the Hungarian data. The published information that has been available since 2014 says that while in 2014, 41,0% had to face social or financial deprivation, in 2018 this was only 20,1%. But even with this significant improvement, we are in one of the worst positions in the EU, only 5 countries behind us, while the EU-28's average is 12,8%. (https://appsso.eurostat.ec.europa.eu/nui/show.do?dataset=ilc_mdsd07&lang=en).

**Figure 1: People at risk of poverty or social exclusion (AROPE)[5][6] (Source: Eurostat)**

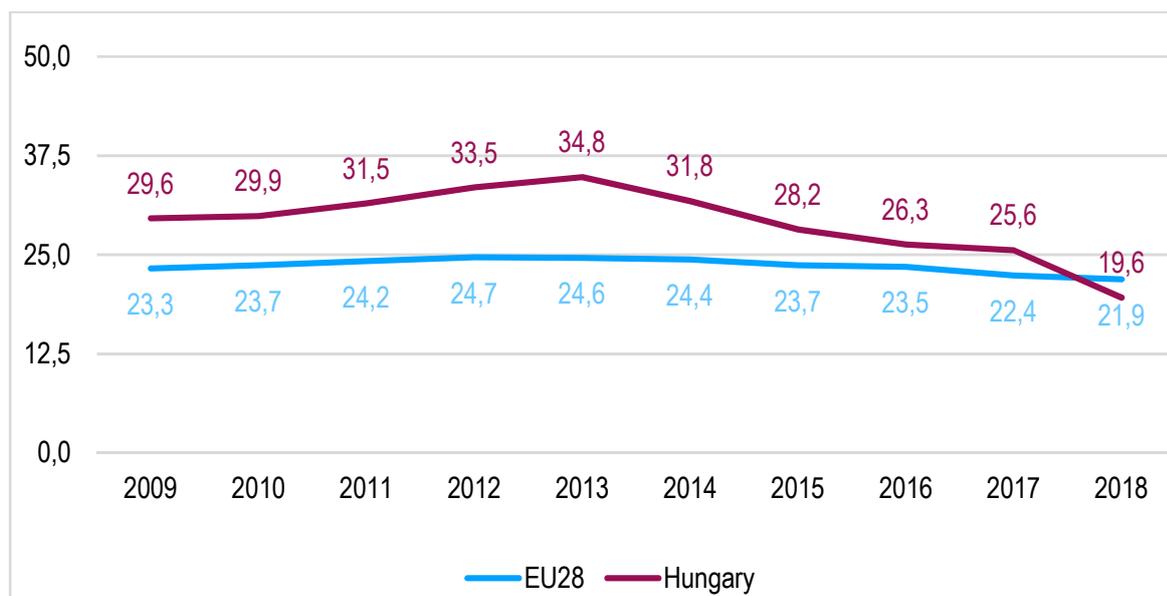

If we look at the most recent data[7], it show that in 2018, 18,9% of the total population, 1.813.000 people were affected by poverty or the risk of social exclusion. If we look at the certain dimensions of AROPE, we see that the dynamics of the dimensions differ: the **relative income poverty** decreases to a lesser extent than the number of severely materially deprived or the number of people living in a household with a very low work intensity. Therefore, the significant decrease in AROPE is explained by the latter two indicators while the relative income poverty is still high but lower than the EU-28 average (16,9%). For this study the risk of poverty is the most important since we examined the households' financial management. Here we can see from data gathered in 2018 that in comparison with the other V4 countries, Poland is in a worse position (14%) while Slovakia (12,2%) and the Czech Republic (9,6%) are in a better position as their indicators of the risk of poverty show.

---

[5] In 2009, Croatian data were not available therefore it refers to the EU-27.
[6] It is important to mention, that the Eurostat data show the year of the collection not the reference period. Therefore, what is mentioned as 2018 by Eurostat, refers to the year of 2017 in the Hungarian statistics (Hungarian Central Statistical Office - KSH). KSH has already published the data for 2018 that were gathered in 2019 which shows that the risk of social exclusion kept decreasing, most recent data show it is 18,9%. The Hungarian data about 2014 refer to year 2015 in case of Eurostat.
[7] http://www.ksh.hu/docs/hun/xftp/idoszaki/hazteletszinv/2018/index.html#chapter-9

**Figure 2: Three different indicators of AROPE (Source: HCSO[8])**

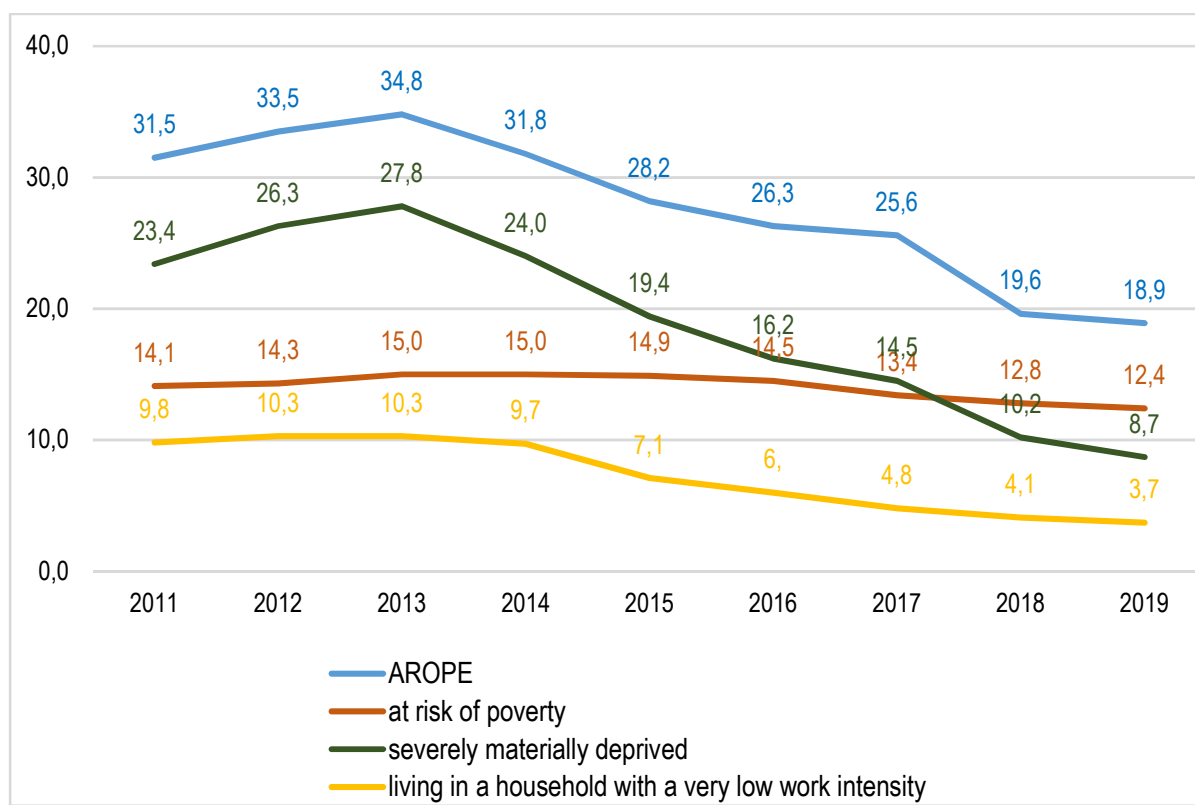

The risk of poverty or social exclusion affects those with no education (33,9%), and single parent households (31,8%) the most. According to the HCSO data, 63,2% of the Roma population is threatened by poverty or social exclusion[9]. We can see, that in the past years, the risk of poverty or social exclusion is significantly decreasing according to the poverty indicators. However, the groups most affected by poverty are the Roma, the parents raising their children alone, and people with low education level. Also, the risk of poverty has not changed significantly in the past years. If we take a closer look at the relative median at-risk-of-poverty gap index, the picture is less appealing. This number shows the difference between people under the poverty threshold's equivalent median net income[10] and the poverty threshold presented the difference in the percentage of the poverty threshold. This indicator is based on data from 2018 when it was 28,9% in Hungary and grew significantly in the past years. This is in direct connection with the income growth of the past years but also shows that the median income of people under the poverty threshold is becoming increasingly distant from the poverty threshold.

---

[8] Hungarian Central Statistical Office
[9] Living standards of households, 2018 HCSO, 2019.
[10] Median equivalent income: the middle value of the sorted list calculated by the income of the total population OECD2 per consumption unit, which means the number of people having lower and higher income is the same.

However, the inequality indicators (GINI, S80/S20)11 has not significantly changed in the previous years in Hungary, so income inequality seems unchanged and positioned on a medium level in comparison with other EU countries.[12][13]

In this paper, I can highlight only a few of the numerous aspects discussed in the literature concerning financial management of people living in poverty. In this case, I focus on 3 concepts: embeddedness, social capital and informality. The notion of embeddedness is connected to the work of Mark Granovetter (1985). According to him, all actions are socially determined and therefore cannot be solely explained by the motivations of the individual and additionally, institutes are not shaped automatically but are constructed socially. Consequently, all economic actions are embedded in a network of interpersonal connections and economic institutions are socially constructed by society. In his opinion, these statements are not compatible with the arguments of neoclassical economics (Granovetter 1990).

Economic actions cannot be understood without the relational and the structural aspects of embeddedness. But this does not mean that actions are automatically and unconditionally follow certain norms or habits. In one hand, accepting that, we would assume that human actions are oversocialized. On the other hand, ignoring that, we would consider individual human action undersocialized. I think it is important, and later we will see that the financial managements of poor households are very much influenced by the interpersonal connections and structures when it comes to transactions.

In their study, Portes and Sensenbrenner defined society as follows: „those expectations for action within a collectivity that affect the economic goals and goal-seeking behavior of its members, even if these expectations are not oriented toward the economic sphere" (Portes, Sensenbrenner 1993, 1323.). They distinguished four types of social capital.

The first type is value introjection, and means that the economic transactions have a moral element that one acquires through socialization, and are based on consensus of the community. The consensus is originated in unwritten rules that one follows because of the consensus of values of the community. Thus, formal regulation is not needed, the community sanctions breaking these rules. The second type of capital is reciprocity exchanges, a system of mutual favors. It is not based on higher moral grounds but on one's balanced social transactions. Not balancing the transactions would result in exclusion from the community.

---

[11] GINI: a possible aggregate statistic indicator of income-concentration between 0 and 1. If the Gini-coefficient is 0, everyone has the same amount of income, meaning total equality. But if it shows 1, then all income is gathered by one person, then it is absolute inequality.; S80/S20 (income quint rate): the rate of the average income of the top and the bottom 20% by the equivalent income dispersion.

[12] See inequality indicators in international comparison:
https://appsso.eurostat.ec.europa.eu/nui/show.do?dataset=ilc_di12c&lang=en
https://ec.europa.eu/eurostat/tgm/table.do?tab=table&init=1&language=en&pcode=tessi180&plugin=1

[13] Let me note, that for measuring poverty there are not only relative but absolute numbers as well. The international literature usually defines such income below 1 or 2 USD per day (Banerjee - Duflo 2007).

The third social capital-type is the bounded solidarity, which is originated in shared hardships and solidarity with other members of the community having conflicts with external groups. It is a sort group reaction to external conflicts, strengthening the group and providing security for the individual.

Finally, the fourth social capital is the enforceable trust. In this case the individual subordinates their current needs and aspirations to the collective group expectations and needs, in the hope of future advantages that may ensure group membership. While bounded solidarity is more value-oriented and often motivated by defense against external elements, the enforced trust is more guided by instrumentalized considerations and is forced by sanctions within the group. It is important to note that social capital sources also have costs and not only positive effects. One cost is the problem of free-riders when strong solidarity networks force the individual in better position to share their assets with those members in need of the community.

We can also consider a cost when strong community networks and norms interfere with the liberty of the individual 's attempt to maintaining external relationships. These two effects can be described as marginal costs that effect and hold mobility back directly. „The mechanism at work is the fear that a solidarity born out of common adversity would be undermined by the departure of the more successful members. Each success story saps the morale of a group, if that morale is built precisely on the limited possibilities for ascent under an oppressive social order." (Portes, Sensenbrenner 1993, 1342). Later I will give examples of these social capitals are present in the financial management practices of the poor.

Arthur Lewis was the first to coin the term of informal sector when examining developing countries, he saw, that as a consequence of the decreasing labor-need in agriculture, manpower became available to an extent that the industry sector could not absorb.

He presumed that this labor-surplus will be absorbed by the formal labor market making it disappear. We have then observed, that the informal economy is since thriving and, in some way, became part of the formal sector and plays a significant role in the subsistence strategies of the poor (Becker 2004). In the financial management of those living in poverty, informality outside the formal institutions plays an immense part. Informality in Central and Eastern Europe has a vast literature mostly in the context of how informal economic activities appear in subsistence strategies and how those impacted beyond the official economy of the state (Danyi-Vigvári 2019).

The role of ethnical dimension in poverty is disputed (Sárkány 2010). Can we talk about poverty only or does the ethnicization of the issue by the dominant discourse polarize the question as in the case of Hungary poverty significantly affect the Roma (Brazzabeni, Cunha and Fotta 2016)? Research shows strong relation between the embeddedness of poor households' management strategies (Bouman 1990), and the regular emergence of financial crises (Collins at al, 2009). Poor households often mutually support one another, and they tend to diversify their activities (Banerjee & Duflo 2011).

Many researches and publications were published in Hungary, that examined the subsistence strategies and economic situations of the poor and/or the Roma population (Dupcsik 1997; Szuhay 1999; Durst 2002; Virág; 2010; Messing & Molnár, 2011; Kotics 2012; Cseri & Orsós 2013). Gosztonyi (2017) - after his research in a village - described the characteristics of loaning practices of families living in extreme poverty: the author highlighted the importance of informal loans, which are often indispensable to financially balance the households. Gosztonyi illustrated the attempts for the financial stability of the households with metaphor of constant juggling. The informal systems, and the network of relatives have a particularly big importance both economically and from an identification point of view within the (Roma) households in extreme poverty (Kovai 2017). Berta (2010) highlighted the importance of conspicuous consumption when a seemingly irrational transaction can play an important part within a community in the construction of identity, in sustaining status, while the majority society is stigmatizing the attitude with a moralizing undertone, presenting the behavior as examples of bad management and wastage.

**Research questions**

This study is trying to answer the following question: how does a poor, Hungarian households' income and expense structure look like today? How often do they face financial problems, and what external resources can they activate to help themselves in these situations? How often do they have overdue debts, loans, and where do these come from? Finally, I try to draw some conclusions about the financial culture of the households in extreme poverty primarily based on qualitative research.

**Methodology - databases**

This study is empirically based on two quantitative surveys, and additional qualitative interviews and field experiences. One of the surveys was requested by the Hungarian Charity Service of the Order of Malta - HCSOM (Magyar Máltai Szeretetszolgálat) where - using representative models - households' financial conditions, health status, access to public services, and living conditions in 13 different towns (either villages or cities) were examined. 1108 household were approached, the data was collected during the fall of 2018.[14]

The other survey was requested by Corvinus University of University[15] where households representing towns below city status in the county of Borsod-Abaúj-Zemplén were examined regarding the aforementioned households' financial management and subjective health status. 508 household were asked, 57 small town in Borsod-Abaúj-Zemplén county, the data collection was done in the spring of 2019.[16] Hereinafter we refer to the first database as HCSOM while to the other database as BAZ.

The two surveys are similar, they were both collected in disadvantaged regions and settlements. While the data collection by HCSOM was done in especially disadvantageous settlements, the BAZ data collection was done is towns in better position, but Borsod-Abaúj-Zemplén county itself is generally disadvantaged. Therefore, it is safe to say, that even though the methods were not similar, both of the collections were focusing on individuals living in extreme poverty.[17] Validating the survey data is obtainable through KSH's Household finance and Living conditions data gathering. [18] In the data of KSH, the lowest segment's monthly income per capita is 35.700 HUF and that of the second lowest segment is 60.300 HUF while the third lowest income is 74.200 HUF in 2018.

In the HCSOM data, the average income per capita in a family, as I talk about it in details later, is 44.200 HUF while according to the data gathered in BAZ, this income per capita is 77.000 HUF. This means the participants in the HCSOM data gathering are positioned somewhere between the two groups with the lowest income while people in BAZ are in the three lowest categories. This seems to be confirmed by HCSOM where the equivalent median income is 98.400 HUF, while in BAZ, it is 117.700 HUF. As I mentioned earlier, the poverty line (60% of the equivalent median income) is 93.313 HUF which means the HCSOM results are very close to this and 46% of the median income per consumption unit is below the poverty line while in BAZ, as seen in the average income segment, it shows a slightly higher rate.

---

[14] The data collection was done by Soreco Társadalomkutatási Tanácsadó Bt. based on records of personal questioning. The questionnaires were filled in in 13 different settlements where the HCSOM is running development programs. Details about these programs are available here: https://jelenlet.maltai.hu/#. The settlements where the following, with the number of households in parenthesis: Abádszalók (75) , Erk (100), Gyulaj (96), Kunhegyes (76), Miskolc-Lyukóvölgy (100), Prügy (107), Ózd-Sajóvárkony (100), Tarnabod (101), Tiszabő (101), Tiszabura (102), Tiszagyenda (50), Tiszaroff (50), Tomajmonostora (50). The data-gathering was part of the EFOP-5.2.4-17 "Társadalmi innovációk – Adaptációk, új módszerek kiterjesztése', a VELUX FOUNDATIONS – From conception to employment", and the EFOP-1.5.1-17-2017-0002 "Segítő jelenlét születéstől a szakképzésig a Kunhegyesi járásban" project.
[15] The research was supported by the Higher Education Institutional Excellence Program of the Ministry of Innovation and Technology in the framework of the 'Financial and Public Services' research project (NKFIH-1163-10/2019) at Corvinus University of Budapest.
[16] The survey was done by Soreco Research Kutató és Elemző Kft. based on  personal questioning.
[17] Poverty itself is hard to define or can only be identified by certain measurements but the exact distribution is unknown in a way that it can be identified on a town level. Therefore creating a classical representative sample about poor people without any kind of measuring number can hardly be done. Databases used here does not represent people living in extreme poverty country-wide but satisfactorily unfolds and shows internal connections and - within appropriate limitations - some characteristics.
[18] This survey is focused on the households' living conditions and consumption every year. The data gathering includes 9500 households asked by the KSH where they question every member of a household including over the age of 16.

The qualitative empirical content is based on 14 interviews recorded in small towns in Borsod-Abaúj-Zemplén county during the summer of 2018 focusing the financial management of the households.[19].

**Results**

**Financial Structures of the households**

In the HCSOM sample, the households were thoroughly questioned about their net incomes from different sources, covered in the previous month. The results are displayed in the chart below (Table 1). In the first column, the individual income sources are listed: labor income and social transfer.

The second column shows the percentage of the households receiving income from the given type of sources. This shows, that nearly 80% of the households receive some sort of labor-based income (labor income, public work, occasional work) and 84% of the households receives some type of social transfer income (pension, childcare allowance, family support, other types of aids). If we take pension out from this analysis and only consider childcare allowance, family support, other types of aids, still 72% receives one or more types of transfer income. It is also clear, that investment income (enterprise, land leasing, apartment rental) is only available for a minimal number of households. 12% of the households needed some sort of financial help from family or loan by the end of the month.

The third column shows the total average income of the households regardless of the number of people living in the household.[20] We see that the average job salary is a net 170.000 HUF, the average public work income is 80.000 HUF, and the average occasional work payment is 70.000 HUF. The aids and supports out of the social transfers make up to an average of 20.500 HUF. Those requesting financial help from family, usually ask for 24.000 HUF, while the loans (presumably from banks or non-relative loans) ask for more than double of this.

---

[19] These interviews were also part of the research supported by the Higher Education Institutional Excellence Program of the Ministry of Innovation and Technology in the framework of the 'Financial and Public Services' research project (NKFIH-1163-10/2019) at Corvinus University of Budapest.

[20] Those households that receives the certain type income.

The fourth column shows the structure of total revenues per capita and per consumption unit[21] compared to the average households. 38% of the income comes from labor income, and slightly more than 13-12% from public and occasional work. Therefore, if we look at public work as a social transfer and not as a regular income, it is safe to say, that half of the households' income is based on a social transfer. Finally, in the last column I showed the total amount of income per one person and per consumption unit.

Adding up either forms of income, the total income per capita is 47.300 HUF (per consumption unit: 105.400 HUF). Besides the previous month's detailed income, the questionnaire also asked about the households' average monthly budget which was 44.200 HUF (98.400 HUF per consumption unit), the difference between the two does not seem substantial. Detailed data show a higher number than the total income, but it is not the peculiarity of this the data gathering, other researchers reported of similar experience (Havasi-Altorjai 2005).

---

[21] Whenever we mention consumption unit, we mean the OECD2 calculated data

**Table 1: Income structure of households according to the HCSOM sample**

| | Ratio of receiving households | Average income of households those receiving a certain type of income HUF) | Structure of total income per consumption unit (all households) | Total income per consumption unit (all households HUF) |
|---|---|---|---|---|
| **Labor income (steady employment that is not public work)** | 44% | *168.900* | 40% | *42.200* |
| **Income from public work** | 33% | 76.900 | 12% | 17.400 |
| **Occasional or seasonal income** | 34% | *69.400* | 11% | *14.800* |
| **Investment income from enterprise, land leasing, apartment leasing, etc.** | 1% | *119.400* | 1% | *41.800* |
| **Disability pension, pension** | 21% | 81.700 | 16% | 27.700 |
| **Childcare allowance, maternity leave** | 31% | 35.600 | 4% | 6.600 |
| **Family allowance** | 66% | 37.900 | 8% | 7.700 |
| **Vocational training scholarship** | 3% | 27.000 | 0% | 5.100 |
| **Supports, benefits[22]** | 27% | 20.500 | 3% | 4.500 |
| **Family support** | 6% | 24.000 | 1% | 6.500 |
| **Loan** | 7% | 51.100 | 2% | 10.400 |
| **Other income[23]** | 7% | 65.300 | 3% | 15.700 |

---

[22] We list support here as follows: support substituting employment, unemployment benefits, regular childcare allowance, child protection allowance, nursing allowance and household maintenance support, and other occasional municipal support.

[23] In the other income category we received answers as follows: allowances for orphans, childcare allowance, income from selling small quantity of product, foster parenting fee.

**Structure of household expenditures**

I thoroughly examined the structures of expenditures of households on the sample of the HCSOM. Among the typical monthly expenses (food, utilities, etc.), there are other, typically occasionally expenses (for example clothing). In such cases, the interviewee gave an estimate amount broken down to a monthly average. In the first column, the different types of expenses are shown. In the second column, the ratio is the percentage of household with such type of expense. In this, it is clearly visible that a huge percentage (78%) has medical, tobacco, or alcohol consumption related expense (63%). 18% of the households has some sort of loan repayment, 5% arrear related expense, and almost every tenth household has some repayment to a private person.

The third column contains the total expenditures of the households independently from the number of persons living there. In the fourth column, I show the ratio of the expenditure structure per capita. In this, it is clearly visible that half of the expenses consists of food and luxury items (alcohol, tobacco, etc.).[24] The Hungarian Central Statistical Office's (HCSO) Household Budget and Living Conditions database gives an opportunity to compare the results with national data. According to this, nationally the households spend far less (around the quarter of their income) on food. Lower income households seem to spend more on tobacco and alcohol: around 8% of their total income while only 3,3% of the households nationwide. Besides those mentioned, 7% is spent on clothes compared to the 4,2% nationwide[25]. For utilities, they spend 16% of their income (the national data is 19,3%), on internet, TV and mobile services the poorer households spend 8% of their income. 6% spent on traveling which is almost half of the national average. The medical expenses are slightly larger (6%) compared to the national percentage of 5,1. 5% of their income is spent on loan repayment, utilities arrears, and private loan repayments. If we observe the total amount of expenses per capita, it is 36.200 HUF. Besides the detailed list of expenses in the previous month, the net thousands of forints usually spent was 38.700 HUF per capita, therefore there is no significant difference between the two calculation. It is interesting to note that in case of income, in comparison to the detailed and global data shows the detailed numbers averages are higher, while in case of the expenses the global result shows the higher number.

---

[24] This correlates with the results of Banerjee - Duflo (2007) who examined the financial management in 13 countries. They positioned income spent on food somewhere above 50% but it reached 70% in some countries.
[25] The higher amount spent on clothing could be explained by the higher number of children and also by what Balázs (2017) has also observed in his research, that families try providing everything expected by peer groups from children in school. This is clearly connected to attempt avoiding shame and supporting mobility. Therefore this is not about losing prestige but about a form of ambition towards emancipation.

**Table 2: Expense structures of households according to the HCSOM sample**

|  | Ratio of benefiting households | Full expenses of the households (involved) (HUF) | Expense structure per capita, all households | Total amount of every expense per capita (involved) (HUF) | HCSO monthly consumption expense per capita[26] |
|---|---|---|---|---|---|
| **Food** | 99% | 64.800 | 42% | 16.100 | 24,6% |
| **Tobacco, alcohol** | 63% | 19.600 | 8% | 4.400 | 3,4% |
| **Electricity** | 96% | 9.200 | 6% | 2.300 | 19,3% |
| **Gas** | 58% | 9.400 | 5% | 2.400 | |
| **Water** | 78% | 5.200 | 3% | 1.300 | |
| **Garbage disposal** | 66% | 3.300 | 2% | 900 | |
| **TV, internet** | *76%* | 7.500 | 4% | 1.800 | 6,8%[27] |
| **Cellphone** | *73%* | 8.300 | 4% | 2.100 | |
| **Travel, fuel, vehicle** | 47% | 17.800 | 6% | 4.200 | 11,9% |
| **Medicine, doctor** | *78%* | 10.000 | 6% | 2.400 | 5,1% |
| **Clothes (monthly average)** | *75%* | 16.400 | 7% | 3.800 | 4,2% |
| **Loans (banks and products)** | 18% | 27.300 | 3% | 6.800 | |
| **Arrears (utilities, tax)** | 5% | 18.600 | 1% | 4.500 | |
| **Loan (private)** | 9% | 19.200 | 1% | 3.900 | |
| **Other expenses** | 11% | 29.100 | 2% | 7.200 | |

---

[26] Values here are from the Living standards of households 2018 HCSIM 2019" publication.
[27] The 6.8% data refers to news communications, so it includes mobile phones and internet as well.

Besides the monthly expenses, the research also asked about the amount spent on wood fuel during the winter. This costs 110.000 HUF in every household, broken down to 36.000 HUF per capita. The research also shows, that poorer households spend 40% of their total income on the first week of the month. I elaborate on the causes and explanations of this later.

The data show, that even the poorest households - however hardly - can cover their monthly expenses. The balance of incomes and expenses are examined by deciles in the sample where we do not take the help and occasional loans by relatives into account, it is confirmed that the incomes and the expenses are in balance (Figure 3). It raises the question, whether the hectic nature of the income causes financial disturbances for the families. The above figures show that the income of the previous month, and the average income asked globally do not differ significantly, even though the detailed average income might show bigger dissimilarities.

**Figure 3.: The balance of income and expense on the HCSOM sample by income deciles (1000 HUF)**

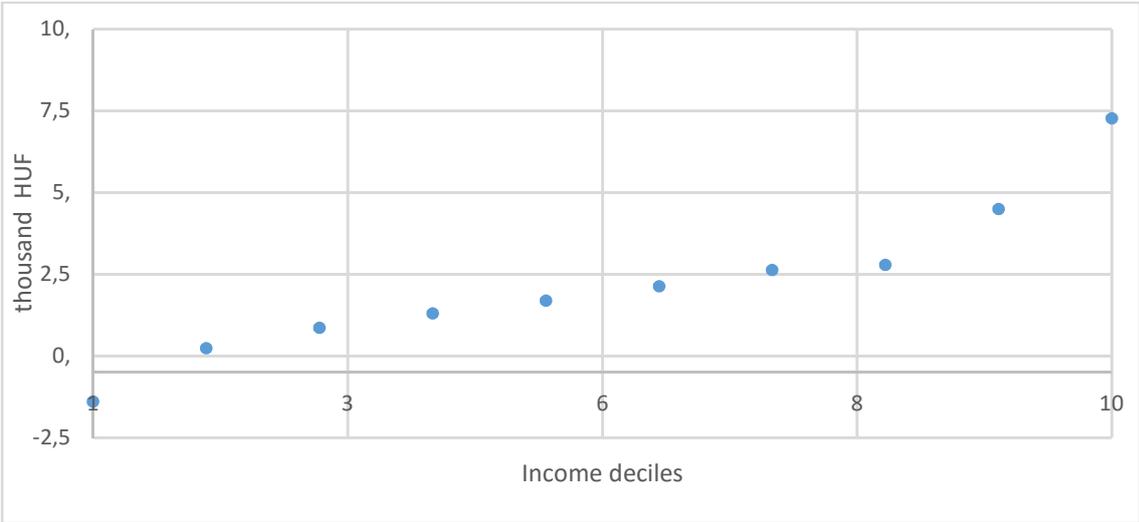

Serious problems occur when families are facing unexpected or bigger expenses. Others pointed out the same events (Mullainathan, Shafir 2013). HCSOM data says that 77% of the households, while the BAZ data shows half of the households encountered unexpected, bigger expenses in the past three years.

**Figure 4: Did you face any bigger or unexpected expenses in the past three years? %**

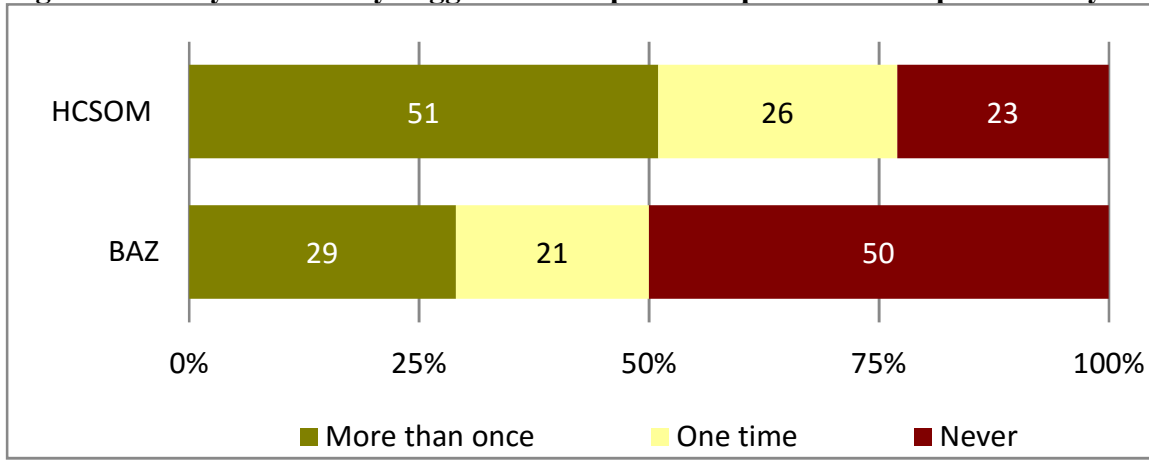

The HCSOM questionnaire also involved questions about the nature of the unexpected expenses. The answers show, that such expenses occurred in relation to schooling, medical reasons, and housing, in most cases. In both databases around the same number of households gave the same answer: 6 out of 10 could cover the irregular, big expenses, but 4 out of 10 had to ask for some kind of loan.[28] So these are the costs that can have a negative effect on the financial balance. This is also confirmed by the data that out of the households having faced more than one of such unexpected expenses, 36% have some kind of arrears (overdue debt, unpaid invoices), whereas among the ones with only one such unexpected expenditures, 24% have arrears. Finally, households with no unexpected expense, "only" 15% have some kind of arrears.

---

[28] From those households that faced such expenses.

**Figure 5: What was the nature of the bigger expense? (if you had such expense), %, HCSOM data**

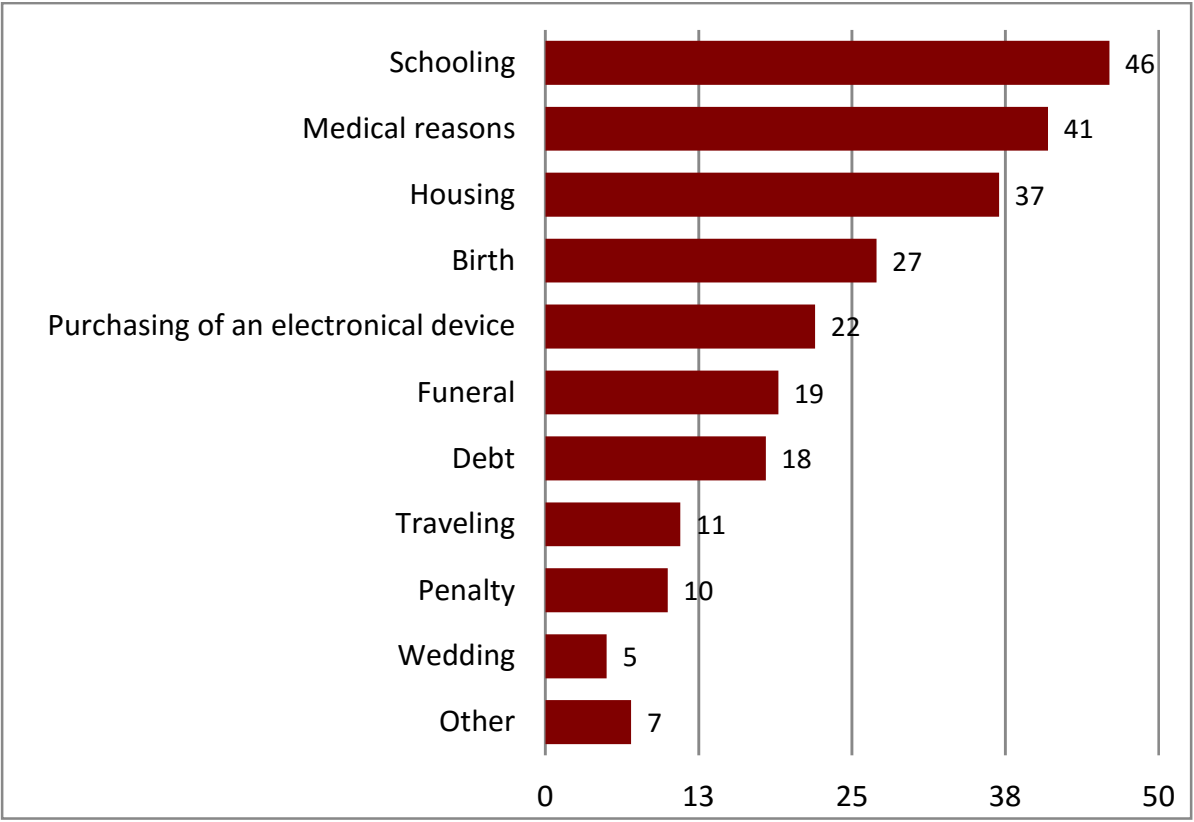

**Figure 6: Were you able to cover the unexpected expenses yourself or did you borrow money?, %**

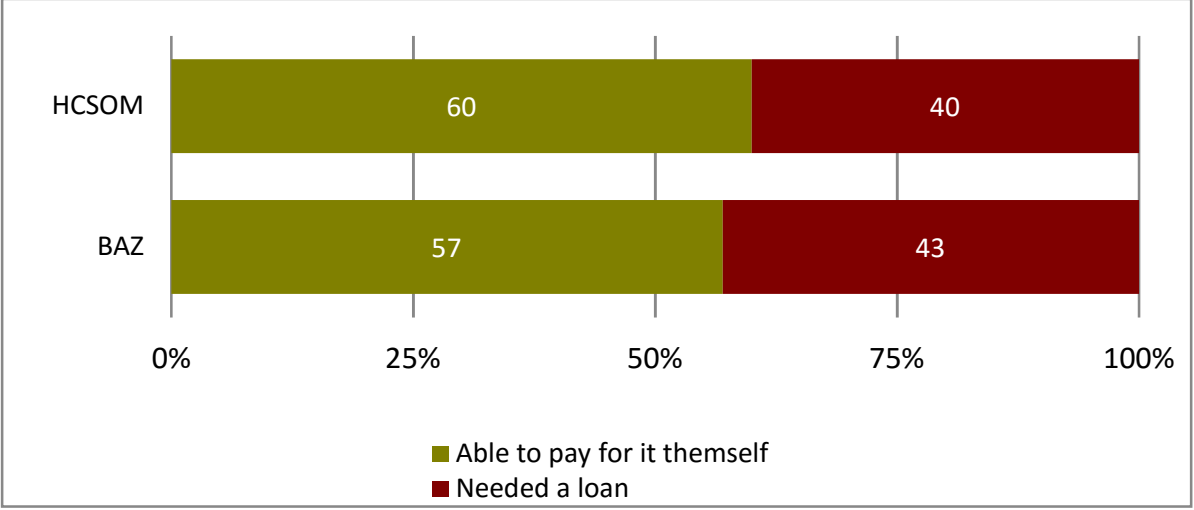

**Household loans and arrears**

Both data gatherings roughly showed the same results: six households out of ten needed to request a loan or other financial help in the past three years. The HCSOM database includes data of families in slightly worse position thus it is especially striking, that only 10% of them said they applied for a bank loan. The table below shows the importance of informal and family and relative based connections.

**Figure 7: Did you ask for financial help from the following in the past three years?, %**

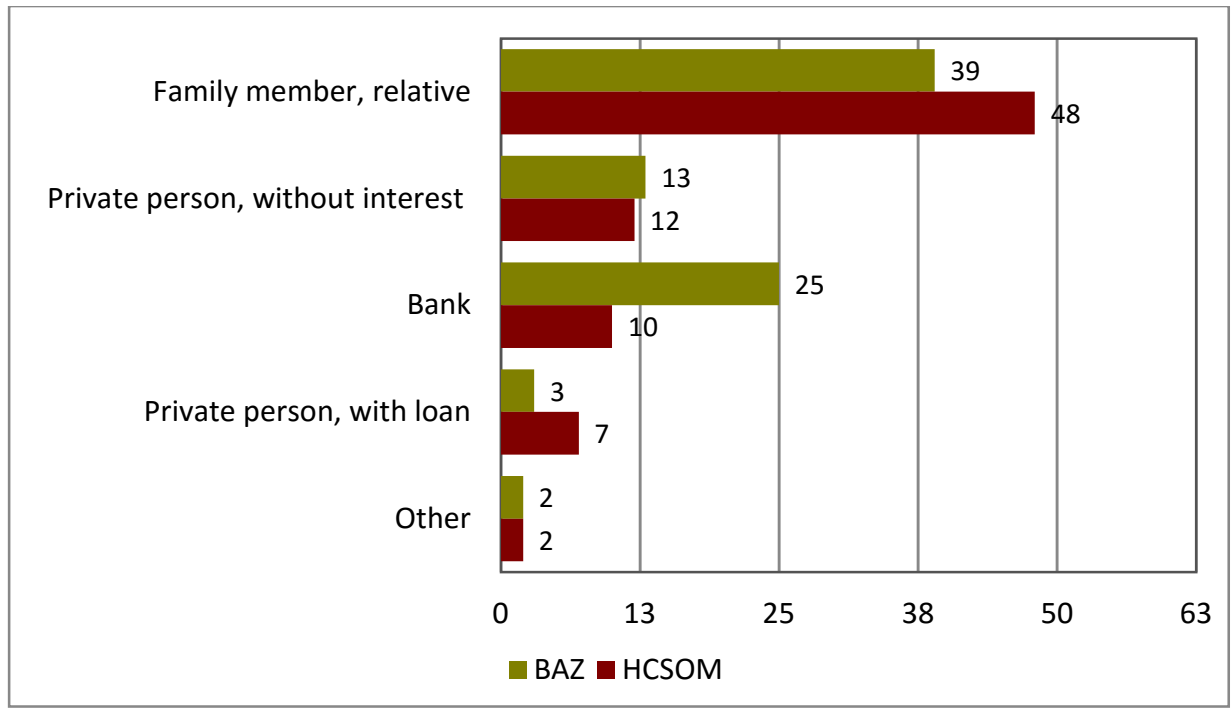

The two databases differ to some extent in terms of the statements of households on how many times they had to ask for financial help at the "end of the month" in the last 12 months. BAZ data shows that in the past year households had to request such help 2,2 times. In these cases, they asked for slightly less, than 20.000 HUF, while HCSOM data shows that in the past year it happened 3,5 times and the amount was around 22.000 HUF.

**Figure 8: How many months were in the last year at the end of which you had to ask for a loan?, %**

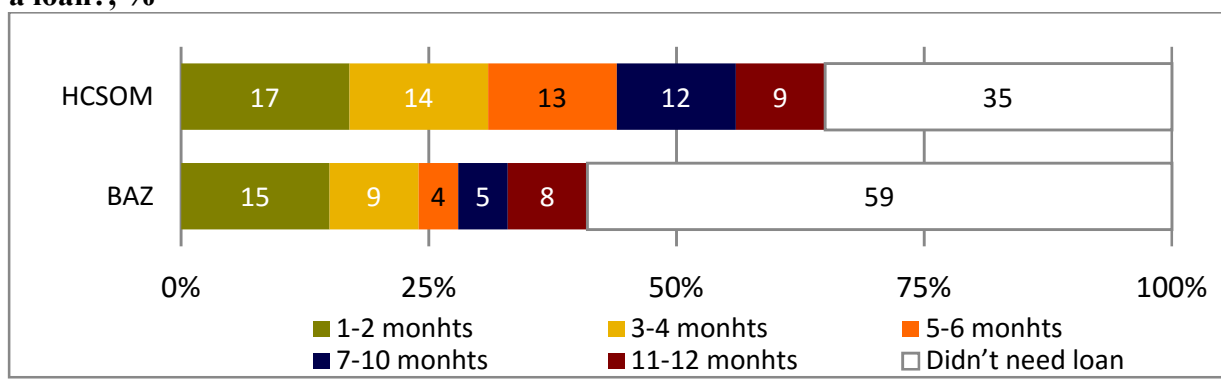

Both surveys asked how much money the household needed to live without serious problems. Basically, one out of ten household replied that they did not need any extra money. The HCSOM sample showed that an average amount of 75.000 HUF, the BAZ sample showed that an average amount of 122.000 HUF would be needed. So, households in a better financial situation mentioned a bigger amount of money they would need to live without any bigger problems. Here, the relative deprivation can be traced, which means people evaluate their conditions in comparison to others, therefore the deprivation is not directly connected to the level of poverty of the individuals.

In a slightly different structure, but both models asked if the household had any debt or arrears. 27% of the households asked by BAZ, and 29% of the households asked by HCSOM reported some kind of debt they could not pay. The HCSOM sample shows, that most of the utility debts for the households occur from unpaid water bills, probably not independently from the fact that shutting off water is the most complicated. The BAZ sample shows a significantly lower number of households in debts to private persons. In this sample we also asked about the amounts of expired debts, which was around an average one million HUF. (Table 3)

**Table 3: Debts and arrears of households**

|  | HCSOM |  | BAZ |
|---|---|---|---|
| **Expired debt, type of arrear** | **Households, %** | **Expired debt, type of arrear** | **Households, %** |
| rent, common charges | 2% | Short term loan | 2% |
| Housing loan | 1% | Housing loan | 5% |
| Short term loan | 2% | Car loan | 1% |
| Other type of missed loan repayment | 4% | Personal loan | 7% |
| Gas check | 2% | Other bank related loan | 4% |
| Electric check | 9% | Utility debt | 13% |
| Water and sewer bill overdue | **11%** | Tax debt | 2% |
| Other utility bill debt | 6% | Debit towards another person | <1% |

| HCSOM | | BAZ | |
|---|---|---|---|
| **Expired debt, type of arrear** | Households, % | **Expired debt, type of arrear** | Households, % |
| Tax debt, missed payment for the National tax and Customs Administration | 2% | Other expired debit | <1% |
| Store debit | 6% | | |
| Debit to family / friends / acquaintances | **11%** | | |
| Other debt | 3% | | |

In the BAZ sample, we examined how the households are connected to formal or informal financial transactions by their income deciles. If we look at the connection between debt and income, we see, that the rate of debt starts decreasing by the 6th decile and this is also true in case of housing (utilities) or informal (to stores or loans from friends) debts (Figure 9). Connection to the formal financial transactions could be measured by checking if the household has a bank account (70%), loan on their house (16%), on their car (5%), whether it is eligible for a bank loan. We see, that in the lower income deciles the number of households having a bank account is linearly growing, and grows above 90% in the 7th decile. Car and housing loans are clearly growing in the $7^{th}$ decile (Figure 10). In case of informal transactions, the exact opposite can be observed. The number of personal loans or expired loans and any other informal financial aids decrease linearly by the income decile (Figure 11). So even households in poverty are quite layered according to their access to the formal financial sphere and how much they have to rely on informal options.

**Figure 9: Household debt by HCSOM sample by income deciles**

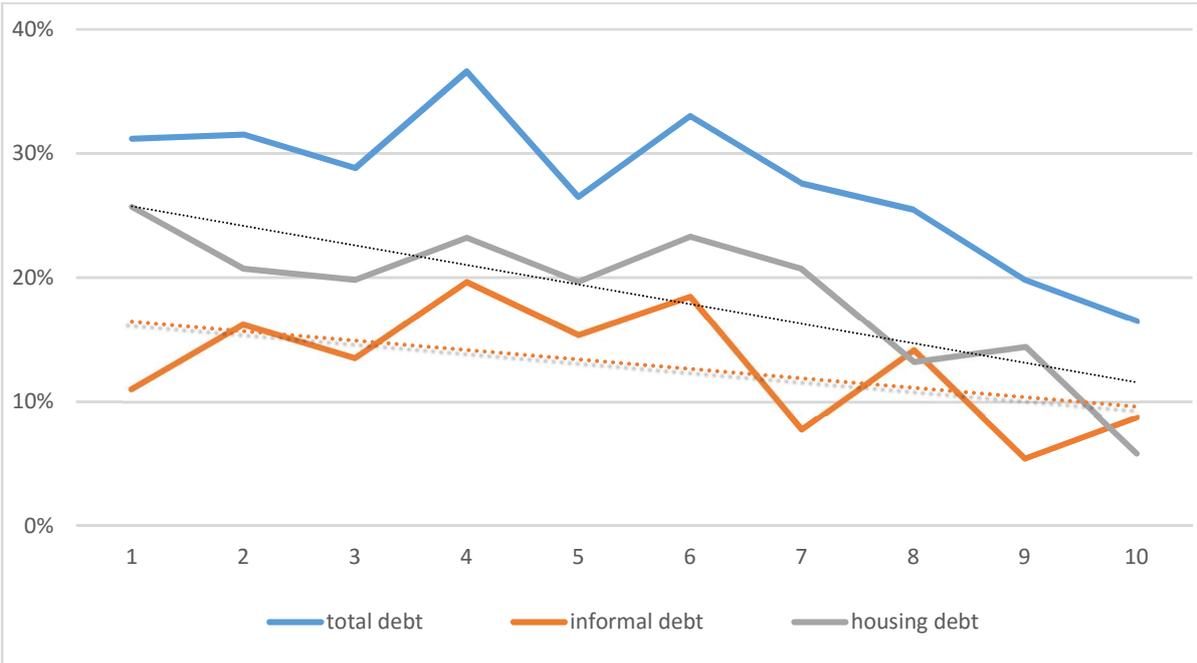

**Figure 10: Car loans, housing loans, and the proportion of those having a bank account by the BAZ sample by income decile**

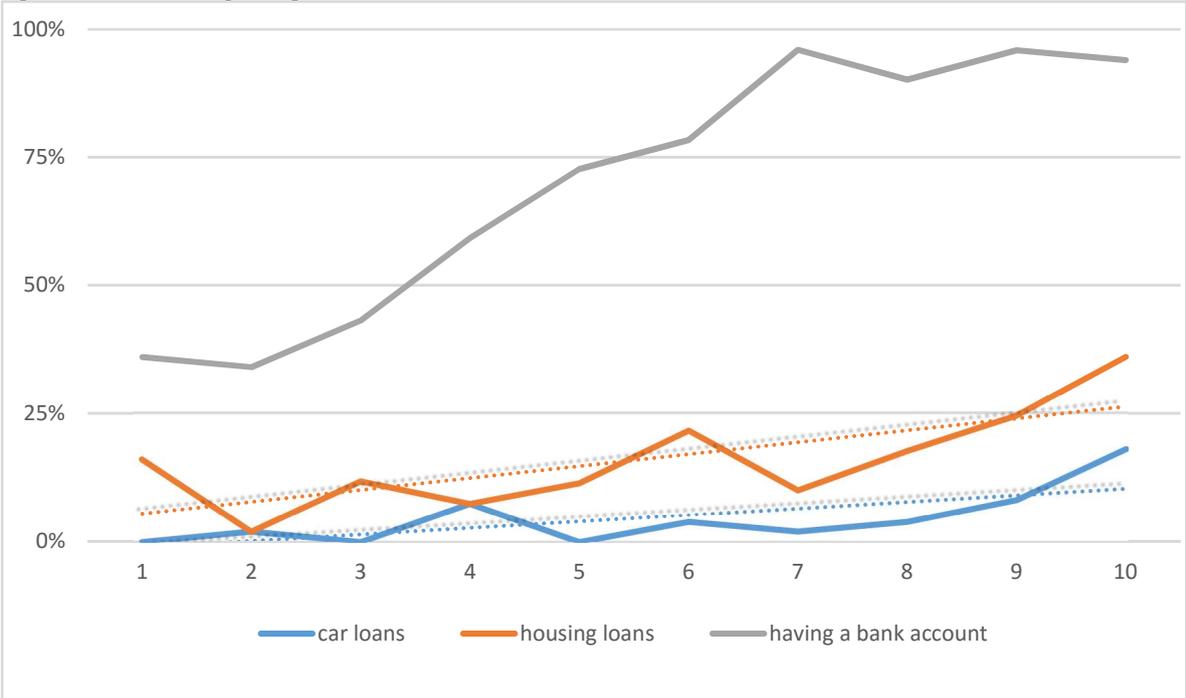

**Figure 11: Rate of households connected to informal financial transactions by the BAZ sample by income decile**

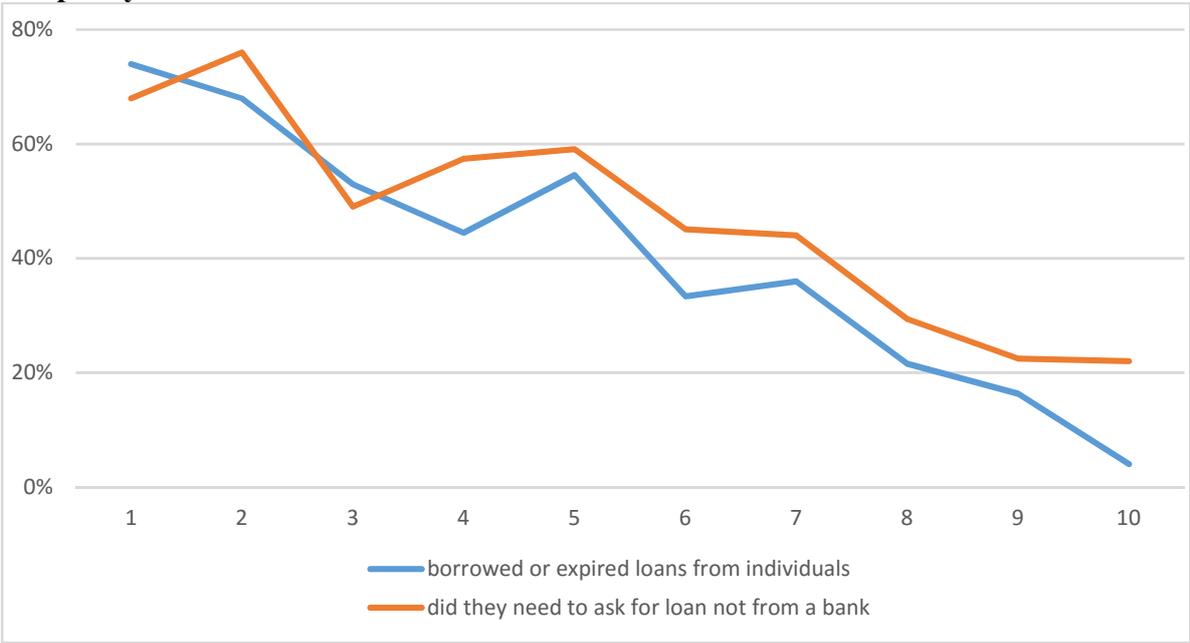

The rate of households having any savings are around 15% in both databases. In the BAZ database we also asked the amount of such savings, their average was around 671.000 HUF.

**Perception of debts**

In the BAZ database we also examined the extent to which respondents are facing disadvantages due to expired loans or whether they knew anyone who did. The perception of loans from relatives or friends are the most common but nearly every second interviewees knew someone who asked for a loan from an individual with interest or someone who did not open a bank account because their debts would have been collected from there.

**Table 4: Do you know someone who has loans?**

| situation | themselves or someone in the household, % | Have met someone elsewhere (also), % | Does not know anyone, %[29] | Weighted scores[30] |
|---|---|---|---|---|
| Had to ask for a loan from relatives, friends | 25 | 55 | 20 | 41 |
| turned off public utilities (water, electricity, gas, etc.) | 15 | 54 | 31 | 31 |
| had an effect on their health | 12 | 47 | 41 | 25 |
| **had to ask for a loan from an individual, with interest** | **10** | **36** | **54** | **20** |
| had to take some of their belonging to the pawnshop | 7 | 44 | 49 | 19 |
| **doesn't open a bank account because their money would be collected from there** | **8** | **38** | **54** | **18** |
| doesn't work officially because their salary would be decreased | 6 | 43 | 51 | 17 |
| had to sell their car | 4 | 45 | 51 | 16 |
| had to move to a smaller or worse (or in worse neighborhood) apartment | 3 | 44 | 53 | 15 |
| had to sell smaller of significant belongings | 2 | 41 | 57 | 13 |
| had to sell their home or it was auctioned. | 2 | 43 | 55 | 13 |

---

[29] This category included those whose households were not affected but had family members or acquaintances, or other people in their settlement affected.
[30] I defined weighted scores as follows: the interviewee was affected =100 points, a member of the household was affected=75 points; a member in the broader family was affected=50 points; a resident of the settlement or an acquaintance was affected =25 points; does not no anyone affected=0 point.

**Transgression within the community**

In the BAZ database, we examined how the interviewees judge the real or imagined transgressions within their community. They had to pick the 5 most severe situations out of 15. I weighted the certain situations according to the number of place respondents gave them: I gave 5 if it was picked for the first place and 1 was given if it was picked for the 5th place. If we look at these real or imagined transgressions (Table 5.) we see that those of related to non-payments are at the end of the list. Of course, these could be partly because these are rather invisible activities unlike those on top of the list. Nevertheless, not paying personal debts is the biggest misdeed while not paying a formal debt is considered a less serious transgression.

**Table 5: Perceived or real transgressions in the community: in your opinion, where you live, what are the five things that people consider the most negative?**

| perceived or real transgressions | Weighted scores | the percentage of people mentioning out (%) |
|---|---|---|
| I someone is using drugs | 1264 | 65 |
| If someone is shouting / quarrelling on the street | 1101 | 68 |
| If someone is littering | 826 | 55 |
| if someone's child doesn't go to school | 825 | 56 |
| If someone is drinking | 561 | 37 |
| If someone is disloyal | 544 | 33 |
| If someone's garden is untidy | 528 | 39 |
| If someone is stealing wood from the forest | 524 | 38 |
| **If someone is not paying personal debts** | **452** | **32** |
| If someone skips public work | 272 | 23 |
| **If someone is not paying utility bills** | **230** | **19** |
| **If someone is not paying off bank loans** | **154** | **10** |
| **If someone is not paying taxes** | **78** | **9** |
| If someone is working unregistered | 76 | 7 |
| If someone is smoking | 62 | 6 |

In conclusion, we see that usually income and expenses of households are in balance, the problem usually comes when unexpected or bigger expenses occur, of which 40% of the households cannot finance by themselves. Data showed that 60% of the households had to ask for loan or financial support in the past 3 years and 30% has overdue loan payment or debt. Even within those living in poverty there is a huge difference between those who can get loans from the formal financial sphere. In the lower deciles it is less likely than within the slightly higher deciles. Not paying debts and loans are not really considered a transgression most probably because it is an emergency and also because it is rather common in the community.

In the following, based on the interviews and the experiences gained in the field, I will highlight some cases to show the practices behind the numbers. I will present the most typical practices through the examples of the 3 households.

**Cases**

<u>First household – buying in bulk, borrowing money from relatives</u>

There are five members in this family. The women is 38, the man is 40, they are life partners with three children. Both of them are public workers, the man has been working there for 5 years, the woman has been working there for 2, neither of them were employed before. The man used to take occasional jobs, mostly at constructions, the woman took seasonal jobs for day labor in a near town. She says she was earning around 3.000 HUF every day while the man earned 7-8.000 HUF every day either undeclared, or sometimes in simplified employment[31]. They live in a house built from social housing subsidy ("szocpol") in 2005[32]. The woman had 6 siblings, the man had 4 siblings, they used to live at the parents' house of the man. From the 3 children, the oldest is a 22-year-old boy, the older girl is 20, and the younger one is 16. The boy and the older girl never finished high school, only the elementary school, they receive Standby Support at the moment. The third child is in high school, the family still receives family support because of her. The family of 5 earns around 170.000 HUF monthly. They spend 30-35.000 HUF on utilities and they usually pay their bills in the beginning of each month. The other high expense is cigarette since all members of the family smokes, costing about 50.000 HUF every month. During the interview, they estimated that the money left after covering

---

[31] Simplified employment is a form of occasional employment, where the employee can work temporarily but only for 90 days in calendar year or for 120 days in case of agricultural seasonal work. More information about the regulation can be found here: 2010. Évi LXXV. törvény az egyszerűsített foglalkoztatásról https://net.jogtar.hu/jogszabaly?docid=a1000075.tv

[32] At that time, after 3 children a 3.2 million HUF non-refundable support was granted to support housing constructions. The full amount of money was only transferred to the family after the completion of the building. Because of this, the new practice for poor families without savings was that a construction contractor built a cheaper house for the family and after the transfer of the Social Policy Allowance ("szocpol"), he took the whole amount including his profit. This system was cancelled in 2009, due to the high number of abuses.

utilities, cigarettes, and occasional expenses is around 80.000 HUF. I would like to show a part of the interview. It is starting with them buying in bulk during the first days of the month in a city nearby:

*Q: And what do you buy and how often?*

*A: Well, every month when we receive it, we pack up.*

*Q: And what is it that you buy then?*

*A: Well, mostly meat products, drumstick and scapula.*

*Q: How much?*

*A: Around 15-20kgs, we freeze it. Sometimes we also buy rump, depends on our options.*

*Q: So, you buy around 20-25kgs meat.*

*A: Yes.*

*Q: And what else?*

*A: Potato, flour, spices, stuff like that.*

*Q: How much potato?*

*A: With a bag of 30kgs. 10kgs of flour and about 6 liters of oil.*

*Q: What else?*

*A: A kilo of Wiener coffee, and a kilo of sugar only for the coffee.*

*Q: Do you smoke?*

*A: Yes, all of us.*

*Q: And do you roll it yourself?*

*A: Yes.*

*Q: How?*

*A: We buy the whole thing at once, around 50.000 goes for it for the whole family.*

*Q: And how many cigarettes do you smoke a day?*

*A: Well, a lot, I smoke about two packages.*

*Q: How many is that?*

*A: Around 40 cigarettes.*

*Q: And your partner?*

*A: He also smokes about the same amount, he works all day. Sometimes I ask the footman.*

*Q: And the girls?*

*A: They also roll it.*

*Q: And how much do you spend for the monthly big shopping?*

*A: Around 50.000.*

*Q: It's enough for how long.*

*A: Well, not for the whole month, about two weeks.*

*Q: What do you do then?*

*A: We wait for the salary.*

*Q: What about bread?*

*A: We pay for it in advance in the store for the whole month, sliced, around 25-30 loafs.*

*Q: Why do you buy bread in advance for a month?*

*A: Because if we spend all of the money, there is no bread. This way it is already paid, and we would always have it.*

The interview highlights several characteristics of the households in extreme poverty. One of these is the phenomenon of monthly shopping in bulk.[33] The families basically invest the money from the income at the beginning of the month. This investment is quite a rational decision for several reasons. Shopping in bigger grocery stores is cheaper than in the small local stores in smaller towns or in villages. But getting to these shopping centers have costs, so it is reasonable to buy bigger amount of goods.[34] Managing and saving money is complicated. Families are continuously exposed to the "drain" of cash. Helping family members and relatives, the constants "requests" of the children are all circumstances that easily cause the "drain" of the scarce family budget easily. If money kept in food, it is much harder to spend and gives the families some sort of security: pasta, flour, potato are always available. The monthly bread quantity paid in advance is a form of such investment and it provides some security.

It is noticeable that the public considers these shopping trips lavish. The perception of the public opinion is that in these regions the shopping centers are "crowded" during the beginning of the month and they spend all the money, they cannot manage their money. I heard many saying that these families cannot take care of money, they spend it as soon as they get it. In reality, these shopping trips should be considered investments, based on quite rational decisions.

---

[33] Balázs (2017) observed contradictory experiences when studied Roma families in the capital and their income opportunities and financial management peculiarities. According to his results, due to weekly payments and the frequency of occasional jobs, families lacked sufficient liquidity to buy in bulk, so their consumption is focused on their urgent needs. I would like to point out, that this is also a matter of perception. Balázs asked shop salespersons who said families might go to stores to buy 1-1 items several times per day. I think this does not necessarily mean these families do not buy in bulk since, as it is clear from the interviews, these never last until the end of the month, therefore at a certain point, occasional shopping is necessary for purchasing items in smaller quantities.

[34] In several settlements, I observed specialized "businesses" focusing on taking a member of the family to the city and back by car to do the shopping. Clearly the cost of this service depends on the distance, but often it reaches the amount of 5000 HUF. The other reason is that many villages lack an ATM and getting money from a post office is expensive. Therefore these trips also connected to the use of credit cards.

It was clearly visible from the research data how often a household needs financial support during a month. We also noticed, how important formal and informal personal connections are among individuals when it comes to financial transactions.

One interviewee described the situation as follows:

> Q: What do you do when you run out of money?
> A: Then I ask my siblings for help.
> Q: How much do you ask for?
> A: Usually 10,000 so we can make it through the month.
> Q: And so, you pay it back right after getting money?
> A: Of course.

These types of transactions are the most common practices of informal loans. These can be called favor loan.

<u>Second household - the lack of a bank account, difficulties of female employment, prioritizing the payment of bills.</u>

Seven people live in the household in the second case. The interviewee, her three children, two grandchildren and her son-in-law. Four years ago, her partner got into a physical atrocity after a conflict and is now in jail, he is about to come home in four months. The girl from the three children living in the household is 22, the older boy is 20, the younger boy is 14 years old. Both boys are still in school. The girl has her own two little daughters and his partner also lives with them. The interviewee is 39, her imprisoned husband is 41 years old. The interviewee has primary school education, and is a public worker and worked in a city nearby in a factory in three shifts through a labor hire company where she earned 150.000 HUF each month. After six months, her fixed-term contract ended, and traveling was very difficult, it took a lot of time. As of now, she is participating in EU project focusing on competency-building. She receives 45.000 HUF for that. Occasionally she works in a grape farm far away from her home, she earns 3.500-4.000 HUF a day. The total income of the family is complete with the family support they receive. She and her daughter live on separate budgets.

Quite often, they try to avoid formal channels of finances because of their debts. For example, it is a problem, that they do not receive their full salary for public work as arrears and debts are deducted from that. There is no country-wide data published about the extent and prevalence of this problem.

*Q: Do you have a bank account?*

*A: No.*

*Q: But you used to have one?*

*A: I used to have one but now I don't. I also have debts and so I don't have one.*

*Q: I assume they money would be collected.*

*A: Yes. Either the 33% of it or 50.000.*

The biggest obstacle for women to get a job in the primary labor market is housekeeping. Additionally, the traditional forms labor division related to gender roles are much more typical in these households. Women do not have their "own" money but they are the ones who manages the family budget. It is usually the men who have their own money that they can spend. This also shows income inequality that is also a relatively under-researched.

*Q: So, what other work opportunities do you have around here?*

*A: There was some hoeing on the fields, gathering mulberry.*

*Q: Did you ever consider getting back to the chocolate factory?*

*A: Well, I wouldn't be able to take care of the children and also, I'd have to buy the travel tickets.*

*Q: But the kids are older now, aren't they?*

*A: But Roma people cook every day. We don't eat food from the previous days.*

It is also noticeable that the different kind of utility bills are prioritized according to if it is easy (electricity), or difficult (water, or sometimes TV) for the service provider to turn off. The ones that cannot get turned off easily and those debts that cannot be enforced easily, are not so important, or not paid at all.

*Q: Do you have TV subscription?*

*A: I do but I can't pay it. We signed for it when my husband was here. They keep sending the invoices. I have a MinDigTV subscription with free installments.*

*Q: For how long do you have it?*

*A: 3 and a half years. They told me they were not going to turn it off.*

*Q: Do you have internet connection?*

*A: No, only on one of the boys' phone.*

Third household - loans

There are nine people living in the third household, the female interviewee with his partner and their seven children (one girl and six boys). All of the kids are in school, the oldest is 15, the youngest is 7 years old. The interviewee is 33, her partner is 49. His partner is a public worker earning 81.000 HUF with tax allowances. They live in their home since 2008, they built it with the support of social housing subsidy ("szocpol"). The interviewee had been on childcare allowance, never worked on the primary labor market. Besides the salary earned by her partner, they receive 112.000 HUF family support and 25.000 childcare allowance, therefore they have 220.000 HUF income each month. The husband used to work in the capital or in the Transdanubia region at constructions, from where he only returned home in every two weeks. He earned about 90.000 HUF for each of those periods. It is not unusual in this family to run out of money for various reasons. They faced difficulties lately related to a funeral, the local government helped that time. The local government has means by law to help families who are in need occasionally, it can also provide loans without interest. They can also advance public workers a part of their salaries. There is no data or research available now on how often local governments use these options but according to our field experiences, quite rarely. Furthermore, in case of smaller villages that lack local taxes, municipalities do not have many options. Although this kind of support would be easily accessible for families facing financial problems.

*Q: Who do you own money to?*

*A: We got an advance from the local government.*

*Q: Do they usually give advances?*

*A: Well, this time they did but usually they don't.*

It is also quite usual that some assets end up as deposit or in a pawn shop to solve the financial difficulties. In the next part of the interview we can see an example, when the help does not come as money but as an object that can be put in a pawn shop for cash.

*A: My mother-in-law lives nearby, she helped us out.*

*Q: How? Do you go over there and ask for this and that? And then she says if she has any money or not?*

*A: Well, she doesn't always have money, but she has some gold jewelry and she usually gives it to me. And then we go to the pawn shop.*

*Q: And so, you ask for it and go to the pawn shop. How much money do you usually get?*

*A: Well last time this happened, we gave in her bracelet and we got 52.000.*

*Q: And how does this work? How long do you have to get it back?*

*A: You have three months to retrieve it.*

*Q: And you have to pay the same amount?*

*A: Of course not, there's an interest.*

*Q: How much?*

> *A: Well, now we put it in for 52, so if I would retrieve it'd be 60.*
>
> *Q: Does it matter if you retrieve it in 1 month or 2 months?*
>
> *A: No.*
>
> *Q: So, you have 3 months, you get 52 and when you retrieve it's 60. Plus the bus ticket. But what if you don't retrieve it?*
>
> *A: That I wouldn't do, she gave it to me with a good heart to help us, I would retrieve it.*

It is also typical that the loan is not actual money but food in advance. This is also some sort of informal loan service that is called purchase for payment or food usury. There is competition in this case as well.

> *A: Well there are those, who sell food, bacon and stuff.*
>
> *Q: Tell us an example. Because we don't know.*
>
> *A: Sometimes they come at the middle of the month by car, they honk into every household and ask if they need anything, then they go buy it and bring it back to them.*
>
> *Q: So first they ask what people need.*
>
> *A: Yes, but this costs double.*
>
> *Q: And when do you have to pay?*
>
> *A: Either when you receive the family support or the salary.*
>
> *Q: You say, more than one person does this. Are the prices any different?*
>
> *A: Yes, because there are those, who know if they sell it too expensive, I'll just go to the other one and get it from there. They know that they have to sell it cheaper than the neighbor, so that people come to them.*
>
> *Q: And when does the shopping start?*
>
> *A: Usually around the middle of the month or at the end of it when they know that the poor Roma people are out of money and then they come.*

We can also find examples when these favor relations are not based on money but in kind. This is partly because in these households forced solidarity makes the lines between personal property and personal sphere less visible (Balázs, 2015)

> *A: Yes, well, if I go in somewhere and there's no bread or something, they say, here you go, let the kids have some bread.*
>
> *Q: How often did this happen?*
>
> *A: Well quite often…*

Based on the interviews and field experiences we saw many different kinds of informal loans among families living in extreme poverty. We tried to show some examples through these cases above. The practices of informal loans are summarized in the diagram below. The classic loan forms are not mentioned here like usury (money with interest) and loan from a store for interest ("boltocskázás")[35]

**Table 6: Types of informal loaning practices**

|  | **with interest** |  |
|---|---|---|
| usury loan |  | shopping – food usury |
|  |  | „boltocskázás" |
| **in cash** ———————————— | | ———————————— **„in kind"** |
| favor (by relative) |  | guarantee – item lending |
| loans from the municipality, or from the employer (payment in advance) | **no interest** | favor food |

The different types of informal loans are not always bad or hurtful since these can fill gaps where formal transactions through banks, or the help of government are missing. Of course, these often make families in need more vulnerable, but it would be hard for them to manage financial difficulties otherwise. It is more relevant if these are the cause of the financial difficulties forcing these households already in need into a dependent relationship or if they help these families. There is no clear answer for this question. Maybe it is safe to say, that the system of the practices of informal loans has become part of the already existing social conditions. Those who give food-usury, usually own a pub or a store in the village. One who gave regular usury earlier can be the current mayor. As a consequence of vulnerability and suppression capital types easily convert to other forms of authority. The main problem is that there are no additional resources in these communities, and the limited amount of goods are being reallocated unequally, thus securing survival conditions are of interest of those, who are profiting from these dependence systems. From the examples above it is clear how social capital sources - mentioned in the introduction - appear. In these households the most common thing is mutual exchange and forced trust. The operationalization and measurement of different types of capitals could be the subject of another study, here we are only able to register their existence. The role of informality is very well proven as well as the hardship of understanding them without their social embeddedness. Without those, they would seem irrational transactions.

---

[35] For more on this: Béres-Lukács (2008), Béres (2015)

Meanwhile, after a deep analysis, we see, that the management of finances of people in poverty is a complex process that requires precise calculations.

**Conclusion and suggestions**

Finally, I will draw up a few conclusions and suggestions based on the experiences of the research.

(1) The **household as a management and risk-taking entity does not have clear, solid boundaries** among those living in extreme poverty. Favor based transactions are quite common therefore risks and difficulties occur frequently, but resources are distributed: they are based in a broader, "forced" solidarity net. This could help during crisis but during prosperity (in case a household's income position is better), it makes securing savings and investing in externally more difficult.

(2) The possible **savings and investments are embedded in informal relationships**, and are being realized outside of the formal institutions. Even though the **resources inside the informal systems may appear as a low threshold "service" for the households lacking resources** and substituting the access to formal institutions, but they are also uncontrolled and unregulated, therefore they can lead to dependencies in certain cases. but because of these are unregulated and uncontrolled, they easily create a depending situation. The poorer the household is the more likely it needs to acquire the missing income **through informal channels and** the more like it does not able to access formal opportunities.

(3) **Incomes and expenses are usually in balance in these families, bigger and unexpected expenses cause the debts.** If a family is in a deep crisis, their stability disappears easily. Even tough, some families have savings, usually in the form of precious metal, mostly in gold. **First of all, these saving can be easily mobilized and has a high prestige value and on the other hand, they are practically outside of the formal saving structures.** It is uncommon to find other kinds of savings since the value of their real estates are very low and thus practically immobile.

(4) The sustainable **forms of taking risks and responsibilities, based on solidarity within the community**, have not been established. Although these forms play an important, often ideological role in alternative-living communities or in economic experiments, in the case of poor households, **the resource surplus is not the base of long-term planning but an opportunity for immediate consumption.** That is why the consumption rituals are important and are strengthening then community. Furthermore, formal tools of investing resource surplus are not known or in often cases, not available. It is worth to invest the surplus back into the community, so in times of hardship the social capital, a poor family can rely on, can be accessed more easily and quickly.

(5) **As a result of the constant presence of scarcity and deficiency, self-rewarding acts and self-representation are very important.** Both meeting immediate needs and the representation of consumption for status have a powerful emotional and identicfication function. Since the boundaries of possessing resources are fluid, the immediate consumption can be seen as a rational attitude despite it is stigmatized and condemned behavior in the eyes of the public. Overconsumption is often a tool of denying status, they represent the desired mobility. Finally we should not forget that at the same time, consumption is a source of happiness, it is a basic human need (Scitovsky 1992).

(6) **Money typically converts to essential assets for one's subsistence but those are harder to mobilize.** The management is not focusing on money but rather on these assets (see: food stocks). This transaction gives the security, so the household's subsistence is not at risk.

(7) **In networks lacking resources, the conversion of the different types of capital is an easier and faster process.** Namely, the economic capital is convertible to political capital, the social capital is convertible to financial capital and so on. The dependent, asymmetric relationships enable these forms of capitals to become centralized, but as **formal and stable institutionalization is quite uncommon**, the possession of certain types of capital is rather limited and volatile. This last statement is more like a hypothesis, proving it by the research data would not be possible but it would most certainly be a very interesting subject for a future research.

Finally, I would like to suggest two more research directions for further consideration.

(1) There is no literature or broad research on the practice of how companies and authorities enforce the collection of loans, taxes, and utility debts and how those practices push households in poverty to seek for solutions outside of formal economy. It would be worth looking into the strategies of financial institutions, public service providers, and the tax authority, and the strategies of households.

(2) There is no literature or research on occasional aiding practices of municipalities after the changes in regulations in the past years, or on their willingness to give financial support for those in need.[36] We have seen, that municipalities can be the first formal institutes that can help families in crisis with providing financial support, payment advances, and interest free loans. Therefore, it will be quite interesting to explore the experiences and practices of institutions.

---

[36] See example here: Kováts (2015).